# Dynamics of thematic information flows

## D.V. Lande, S.M. Braichevskii
Information center ElVisti, Kyiv, Ukraine

*The studies of the dynamics of topical dataflow of new information in the framework of a logistic model were suggested. The condition of topic balance, when the number of publications on all topics is proportional to the information space and time, was presented. General time dependence of the publication intensity in the Internet, devoted to particular topics, was observed; unlike an exponent model, it has a saturation area. Some limitations of a logistic model were identified opening the way for further research.*

### Topics

One of the main features of network information space is the availability of a dynamic segment [1], its content changing with time. Thereby, recently the concept of data flows has become relevant [2-4], they begin to play more and more important role in present-day information technologies. Therefore, to study the dynamics of data flows is definitely important and interesting, particularly because the issue has not been researched enough [5].

During recent decades certain achievements have been made in solving the problem of information obsolescence in the framework of model Barton-Kebler [6], which was developed because of the need to evaluate a real usage term of scientific works and also the approaches of Cole and other authors [7]. Later it turned out that the results achieved (as well as the approaches) could be useful in a wider context of the information technologies. However, the comprehension of the processes of the dynamics of data flows requires somewhat deeper analysis and more sophisticated technique.

Studying the dynamics of thematic data flows of new information in a framework of a logistic model is suggested in this work. Common time dependence of the publication intensity was received; it appeared to correspond to experimental data. Alongside with this, limitations of a logistic model have been identified which in turn opens the ways for further research.

### Available models

As it is well known, a general structure of the Internet network consists of two main parts – static and dynamic.

The whole Internet space can be relatively divided into two constituents – stable and dynamic, they both have very different characteristics from the point of view of a required integration of data flows. In particular, even information obsolescence processes, loss of its actuality in Barton-Kebler model are described with the equation, which consists of two components:

$$m(t) = 1 - ae^{-T} - be^{-2T},$$

where $m(t)$ – share of useful information in a total dataflow through time, the first numerator corresponds to stable resources, the second one - dynamic-new.

A stable constituent of the Internet contains "long-term" information, while a dynamic constituent has constantly updating resources. Some part of this constituent joins the stable one in the course of time, while a greater part of it "disappears" from the Internet or enters the segment of "hidden" web-space, not accessible for users via known information-retrieval systems.

A segment of new information is apparently more vividly dynamic. On the one hand, it has the highest level of updating; on the other hand, huge amount of data is generated and distributed there. In view of our needs, it is this segment which seems to be the best for the research.

Generally speaking, information dynamics in the network is due to many factors, most of them cannot be analyzed. As a reasonable assumption, general character of time dependence of the number of thematic publications in the network is defined with very simple regularities, which allow developing mathematic models.

In the works we are familiar with and which deal with the information obsolescence, Maltus model is used [8] (probably with some modifications similar to super-position of two curves with different parameters). The advantage of the model is that Maltus equation has an exact answer in the form of a very simple and convenient function – exponent; from the point of view of the result interpretation, it looks very disputable. The main problem is that exponent is a monotonous increasing function: it cannot describe the processes, which by nature must have local extremes.

There is no need to prove that news loses its actuality which results in the decrease of the publication number. To get more adequate dependence, we have to refer to more complicated models.

A logistic model appears to be very promising; it was suggested by P. Ferhlust [9] to describe the dynamics of population and by R. Purl [10] – for biological communities; later it was successfully used in numerous researches. The advantage of the model, first of all, is the fact that it combines the simplicity of the task formulation with the possibility to vary the answers with help of a set of parameters which can have more or less transparent physical contents.

**Topic balance**

Let us look at the general picture of the dynamics of thematic data flows, in particular the mechanisms which are typical for a new segment of the Internet.

We presume that the majority of organizations-generators of new information work in a stationary regime, which can be characterized by maximum capacity of information space $N$ (we state that the issue of parameter regularity and their measurement is not considered in this paper). This means that each organization-generator produces information flow which is constant as to the number of signs and messages. It is the amount of information that changes with time, depending on the topic. In other words, the increase in the publication number on one topic results in the decrease in the publication number on other topics; so for each time span $T$ there is:

$$\int_0^T \sum_{i=1}^M n_i(t)dt = NT, \qquad (1)$$

where $n_i(t)$ – the publication number per time unit, and $M$ – the total number of all possible topics. Part $n_i(t)$ is always expected to be zero.

To study the dynamics of a particular thematic information flow, which is described with the density $n_i(t)$, presents the major interest.

It is worth mentioning that when we say "topic" referring to a information flow, it should not be taken very directly. By using this word we mean certain abstraction/generality associated with the activity of information sources. It does have a connection with the events in a real world, but its subjective expression cannot be as simple as it may look from the first sight. For example, the launch of a spacecraft to Mars may cause a number of publications concerning the expediency to relocate budget finance in favor of research.

Hence, it is not always possible to establish the connection between the activity increase of sources-generators and the situation in the environment. That is why we will speak about the appearance of a new topic, bearing in mind a set of factors which cause the increase in the publication number per time unit. The localization of a particular topic in a semantic space and its articulation in communicative mechanisms is a different issue which we do not have an intention to discuss, at least in this paper. We will only state that it may be solved in a wide range of cases. Our major concern is the fact that topics appear and disappear at certain moments (i.e., they lose their actuality and present no interest for people).

It may be theoretically assumed that lots of publications associated with a defined set of topics are interlinked, namely, some publications can be referred to several different topics at the same time. Generally, such "polytopics" is a phenomenon which should not be ignored, however we, to a first approximation, will consider that it does not distort a general picture.

Furthermore, we will think that during its period of actuality a topic fixes a set of mechanisms which result in the increase of the publication number that have some common features. Different topics may raise different data flows; so in this respect they are not interchangeable. On a formal level, let us compare two parameters with a topic as an abstract concept: duration (typical "life time") $\lambda$ and intensity $D$. In the context of this work, we will consider the intensity to be constant. This is a simplified opinion but it is good enough to identify general trends.

The mentioned-above duration does not necessarily coincide with the beginning and the end of an event or a number of events. It characterizes only a certain timespan when a topic actuality is lost. Intensity can be defined as a quantity which characterizes the number of publications caused by a certain topic averaged on interval $\lambda$.

The response of media means, described as quantity $D$, has never been momentary: there is always a certain time delay. To take into account this factor, let us introduce a factor of lateness $\tau$.

Hence, we can suggest the following quality picture of the dynamics of thematic data flows. The generation of data flows has two constituents: background and thematic. A background constituent is defined with numerous factors which are not very much connected with each other, and under certain conditions it may approach (as to thematic classification) noise. But it ensures the publication of relatively stable number of materials based on the principal "Something should be published!"

A new topic causes the process (to be more exact, a set of processes) of re-distribution of network resources as actual stories appear. The scope of background publications decreases, that of thematic ones increases. If the duration of two or more topics intercrosses, then thematic publications begin to redistribute among them, the nature of the re-distribution being defined with the meaning of $\lambda$ and $D$ of each topic. When a topic loses its actuality, associated resources move either to background flows or to other thematic ones.

In this paper we study the second, thematic constituent, and we focus on the dynamics of flows caused by one topic. A definition "interaction" of several topics is a different subject to be researched and it is beyond our task.

We will give only two real data flows and their behavior in the model which will be described then. In the first case (Fig. 1a) the publications, scanned by the system of news monitoring from the Internet according to the topics of illness and quitting a career by a famous political figure, were considered. Before his illness took a turn for the worst, the publications concerning his activities were at a high level. The information about his illness increased the number of publications considerably; it reached the highest saturation level. The information about his giving up a political career decreased the number of publications to the lowest level; final stabilization occurred at this level. Another example is election of a mayor of a big city (Fig. 1b). Before election campaign started there were very few publications about this person in the Internet, which corresponded to a low stable level.  The election and appointment on the position of mayor were accompanied by a great number of publications of both positive and negative nature (an upper level). The process of a mayor's activities after the election is followed by the number of publications which corresponds to an average stabilizing level.

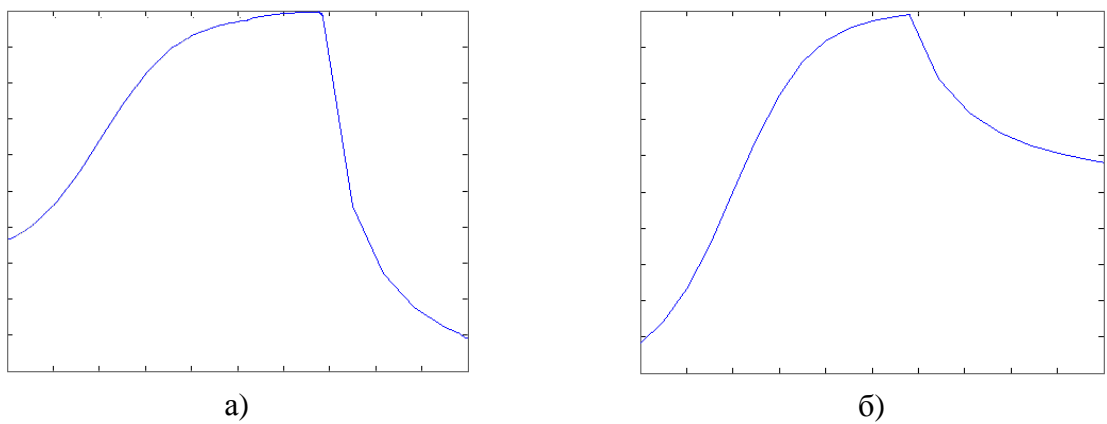

а)  б)

*Fig. 1. Examples of data flows*

**Logistic model**

If necessary, a logistics model can be considered as a generalization of Maltus model, which envisages a balance between the increased speed of a function and its meaning at each time moment:

$$\frac{dn(t)}{dt} = kn(t), \qquad (2)$$

where $k$ – some coefficient of proportion. We agreed to consider the dynamics of a particular thematic information flow, so we will not write indices for quantities $n_i(t)$, which define a topic.

The idea is to make a coefficient in Maltus equation a time function, and the answer should not exceed a threshold meaning. Various ways can help, but the use of constant is the most popular; in its obvious view, it limits the answer increase. In our case we will use capacity $N$. Then we can present the following

$$k(N - rn(t)), \qquad (3)$$

where $k$ – Maltus coefficient, and $r$ – the factor, which describes negative processes for this system, associated with inner factors.

We have to take into account the parameters in obvious view, which characterize the effect of a topic on the publication dynamics.

As the intensity $D$ is defined as constant, its contribution will be represented as follows:

$$y(t) = \begin{cases} D, & 0 < t \leq \lambda \\ 0, & t < 0, t > \lambda \end{cases} \qquad (4)$$

Correspondingly, we will consider two time areas separately: $0 < t \leq \lambda$ з $D > 0$ and $t > \lambda$ with $D = 0$, answers for them are functions $u(t)$ and $v(t)$. We will receive a complete answer by means of "felling" on a boundary in point $\lambda$:

$$n(t) = \begin{cases} u(t), & 0 < t \leq \lambda \\ v(t), & t > \lambda \end{cases}$$

$$u(\lambda) = v(\lambda)$$

(5)

The increase in the publication number on a given topic when its actuality is not equal to zero ($D > 0$), and probably the transfer to a saturation level, corresponds to the first area; the process of decreasing the publication number, caused by the loss of its actuality ($D = 0$), corresponds to the second area. Having adjusted the parameters to the threshold quantity $N$, the equation for the first area will be the following:

$$\frac{du(t-\tau)}{dt} = pu(t-\tau)(1-qu(t-\tau)) + Du(t-\tau),$$

$$u(0) = n_0$$

(6)

Quantity $p$ defines standardized probability for a publication to appear per time unit despite the actuality of a given topic. Such factor shows background mechanisms of the information generation (a typical example is: re-publication of the materials which have previously been published in prestigious information resources). Quantity $D$ characterizes a direct effect of the actuality of a given topic. Parameter $q$ characterizes a speed decrease in the publication number and is the quantity which is inverse to asymptotic meaning of the dependence $u(t)$ when $D = 0$.

The initial condition in (6) expresses two aspects of the information dynamics: firstly, the availability of background constituent of data flows, secondly, uncertainty of a definite moment, when a particular topic contributes to a process of publication generation. Due to this, at a time moment $t = 0$ there exists a certain quantity of publications which can be associated with this topic.

For the second area we have

$$\frac{dv(t-\lambda)}{dt} = pv(t-\lambda)(1-qv(t-\lambda)),$$

$$v(\lambda) = u(\lambda)$$

(7)

Since in the second area a topic has no effect on the publication dynamics (it describes the processes which are inertial to the topic), we do not include factor of delay $\tau$ in equation (6). The threshold condition in equation (7) provides "felling" of functions $u(t)$ and $v(t)$.

The answer (6) is as follows

$$u(t) = \frac{u_s}{1 + (\frac{u_s}{n_0} - 1)\exp[-(p+D)(t-\tau)]},$$

(8)

where $u_s$ – asymptotic meaning $u$, the quantity of which defines the saturation area (if, of course, this dependence has enough time to reach it):

$$u_s = \frac{p+D}{pq}.$$

(9)

We state that expression (9) does not depend on meaning $n_0$, which proves that initial conditions are not important for the saturation of information dynamics. No matter what an initial number of publications is, the saturation will be defined exclusively by the parameters, which characterize background speed of the increase in the publication number, quantitative level of actuality and negative factors of the process. From the practical point of view we may ignore background factors which are not easy to be studied.

Curve (8) has a bending point

$$t_{inf} = \frac{1}{p+D}\ln(\frac{u_s}{n_0}-1) + \tau. \tag{10}$$

Thus, we have so-called S-like dependence for the first area, and when $t \sim t_{inf}$ dependence (8) moves to linear and corresponds to a linear model.

For better convenience we represent (8) in a different way:

$$\frac{u_s \exp[(p+D)(t-\tau)]}{\exp[(p+D)(t-\tau)] + (\frac{u_s}{n_0}-1)} = \frac{u_s \exp[(p+D)t]}{\exp[(p+D)t] + (\frac{u_s}{n_0}-1)\exp[(p+D)\tau]}. \tag{11}$$

It is clearly seen, provided

$$t < \frac{1}{p+D}\ln(\frac{u_s}{n_0}-1) + \tau = t_{inf}. \tag{12}$$

dependence $u(t)$ has exponent nature, its expression being defined with the delay quantity $\tau$. Hence, for meanings $t$, which are much lower than those of $t_{inf}$, our model agrees with an exponent model.

A typical dependence is shown in Fig. 2.

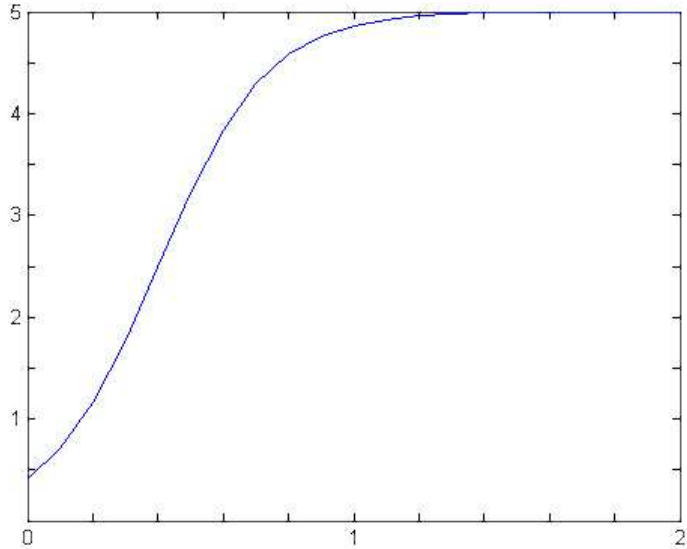

*Fig. 2. Increase area*

Let us move to the second area. Its answer looks as follows:

$$v(t) = \frac{u(\lambda)}{qu(\lambda) + (1-qu(\lambda))\exp[-p(t-\lambda)]}. \tag{13}$$

If dependence $u(t)$ has enough time to reach saturation within timespan $t < \lambda$, we may simplify the answer (13), showing it in the following way:

$$v(t) = \frac{v_s(p+D)}{p + D(1 - \exp[-p(t-\lambda)])}, \tag{14}$$

where $v_s = 1/q$ asymptotic meaning of the dependence $v(t)$.

As it should be expected, quantity $v_s$ depends neither on initial conditions nor on "felling" on area boundaries.

In the second area dynamics of publications to a first approximation has an exponent nature, so it agrees with the results.

A typical dependence of the second area is shown in Fig. 3.

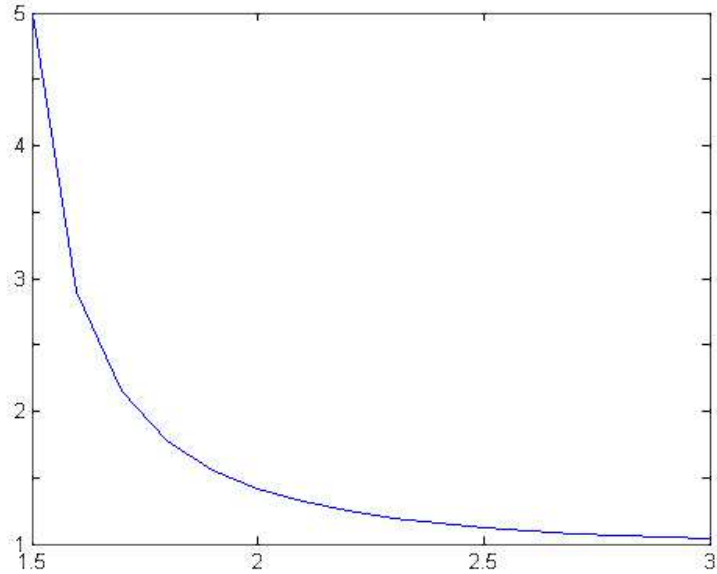

*Рис. 3. Decrease area*

So, we see that our dependence has saturation area $u_s$ (when $t \leq \lambda$) and asymptotes $v_s$, that describes a gradual decrease in the number of publications to a background level. It means that it qualitatively agrees with common opinion about the nature of information dynamics, received from experimental data. Besides, it also agrees with linear and exponent models in certain areas $t$.

A typical complete dependence $n(t)$ is shown in Fig. 4.

**Conclusion**

Thus, a suggested model gives a correct description (at least at a level of qualitative properties) of time dependence of publication density, caused by a particular topic. It contains the saturation area which cannot be explained in the framework of an exponent model.

We also see that the dependence received is not symmetric and has representative "crest" on a boundary of two areas. The answer to our equation for the second area, contrary to the first one, has no saturation condition; it describes closer-to-exponent decline, which asymptotically moves to zero.

This interesting aspect of a curve behavior is practically observed in some cases, but not in all of them. The experiments prove the availability of two more types, which we will not discuss now. We will only mention that the easiest realization of the model has been considered. There is a chance that its more complicated modifications will make it possible to describe all major types of real dynamics.

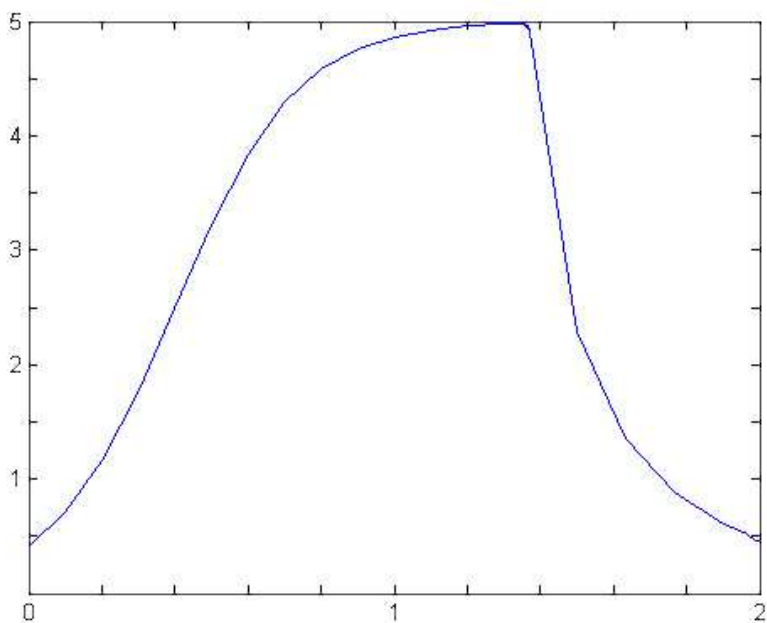

*Fig. 4. Generalized figure of the dynamic of thematic information flow*

Cyclic processes of increase-decrease in the information resource activities present another problem of information dynamics, and they are not directly connected with information factors (for example, periodical decrease in the publication number at weekends/holidays).

The issue of identifying the correlation between the answers of the suggested logistic equations and the balance of topics is open for research (1).

Therefore, we have all grounds to state that a logistic model does describe the dynamic of a certain category of thematic data flows.